\documentclass[aps,pre,twocolumn,superscriptaddress,showkeys]{revtex4-2}

\usepackage{hyperref} 
\usepackage{amsmath}
\usepackage{amssymb}
\usepackage{cleveref}
\usepackage{algorithm} 
\usepackage{algpseudocode} 
\usepackage[normalem]{ulem}
\usepackage{graphicx}
\usepackage{xcolor}

\newcommand{\rev}[1]{\textcolor{black}{#1}}

\begin{document}

\title[SEER]{The energy cost of local rearrangements, not cooperative effects, 
makes a model glass solid}

\author{Massimo Pica Ciamarra}\email{massimo@ntu.edu.sg}
\thanks{These two authors contributed equally}
\affiliation{Division of Physics and Applied Physics, School of Physical and Mathematical Sciences, Nanyang Technological University, 21 Nanyang Link, 637371, Singapore}
\affiliation{Consiglio Nazionale delle Ricerce, CNR-SPIN, Napoli, I-80126, Italy}
\author{Wencheng Ji}
\thanks{These two authors contributed equally}
\affiliation{Department of Physics of Complex Systems, Weizmann Institute of Science, Rehovot, 234 Hertzl St., Israel}
\affiliation{John A. Paulson School of Engineering and Applied Sciences, Harvard University, Cambridge, MA 02138, USA}
\author{Matthieu Wyart} \email{matthieu.wyart@epfl.ch}
\affiliation{Institute of Physics, \'Ecole Polytechnique Fédérale de Lausanne, Lausanne, CH-1015, Switzerland}


\begin{abstract}
Which phenomenon slows down the dynamics in super-cooled liquids and turns them into glasses is a long-standing question of condensed-matter.
Most popular theories posit that 
as the temperature decreases, many events must occur in a coordinated fashion on a growing length scale for relaxation to occur.
Instead, other approaches consider that local barriers associated with the elementary rearrangement of a few particles or `excitations' govern the dynamics.
To resolve this conundrum, our central result is to introduce an algorithm, SEER, which can systematically extract hundreds of excitations and their energy from any given configuration. We also provide a novel measurement of the activation energy, characterizing the liquid dynamics, based on fast quenching and reheating.
We use these two methods in a popular liquid model of polydisperse particles. 
Such polydisperse models are known to capture the hallmarks of the glass transition and can be equilibrated efficiently up to millisecond time scales. 
The analysis reveals that cooperative effects do not control the fragility of such liquids: the change of energy of local barriers determines the change of activation energy. More generally, these methods can now be used to measure the degree of cooperativity of any liquid model.
\end{abstract}


\maketitle

\section*{Introduction}\label{sec1}
Although glass making appeared four thousand years ago \cite{shortland2007trace}, and was already used to make windows in Roman times \cite{fleming1999roman}, the physical process preventing glasses from flowing remains an enduring mystery \cite{Anderson95,Debenedetti01}.
As a liquid is cooled,  the relaxation time $\tau$ below which it acts as a solid continuously grows from picoseconds at high temperatures up to minutes at the glass transition temperature $T_g$.
In liquids called strong such as silica \cite{angell1985strong}, $\tau$ follows an Arrhenius law $\tau=t_0\exp(E_a/T)$,  where $t_0$ is some microscopic time scale, the activation energy $E_a$ is constant and $T$ is the temperature, in units where the Boltzmann constant is unity. 
By contrast, in liquids called fragile, $E_a(T)$ can increase by up to several folds as  $T$ decreases. 
In both cases, as the dynamics slows down it also becomes correlated on a growing length scale $\xi(T)$ \cite{kob1997dynamical,dalle2007spatial}. These phenomena, challenging to explain, occur while the static structure of the material displays limited changes under cooling.

\begin{figure*}[!t]%
\centering
\includegraphics[width=0.8\textwidth]{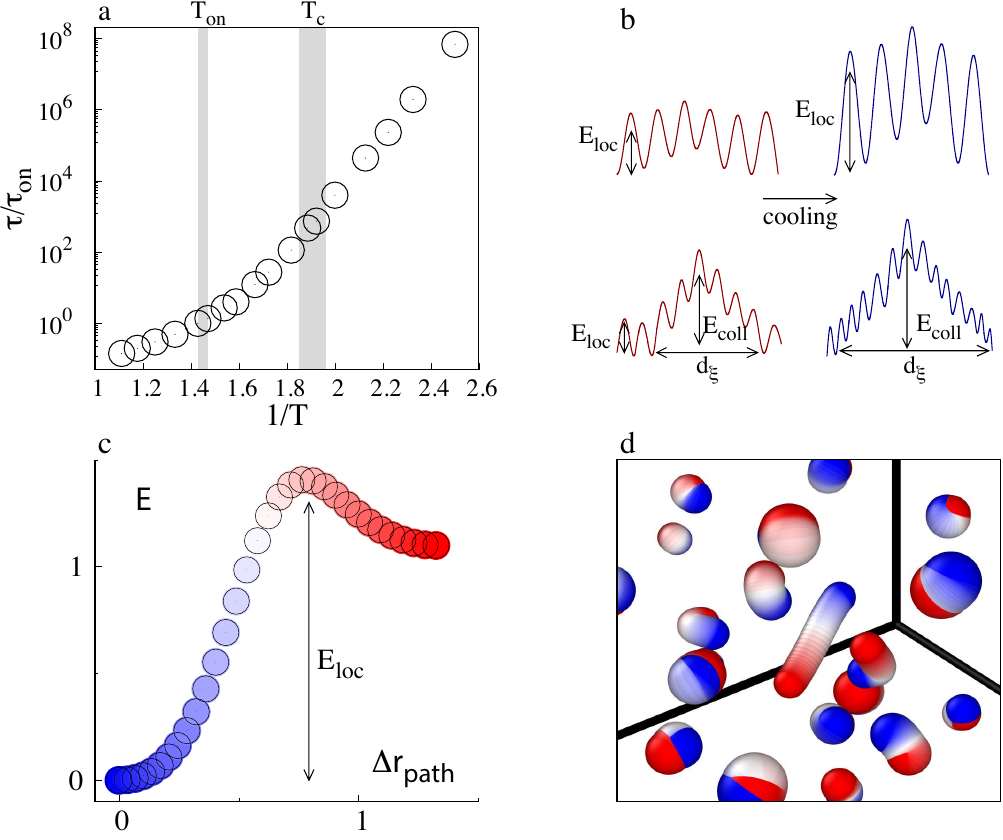}
\caption{
{\bf a} Temperature dependence of the relaxation time $\tau$ estimated from the decay of the overlap function in our poly-disperse model. 
It is customary to define the onset temperature $T_{\rm on}$ as the temperature below which the energy of inherent structures starts to decrease.   $T_{\rm on} \simeq 0.7$~\cite{Lerner19} for the considered model. We denote by $\tau_{\rm on}\equiv \tau(T=T_{\rm on})$ the relaxation time at this temperature. According to mode-coupling theory, the relaxation time diverges as $\tau \propto (T-T_c)^{-\gamma}$. 
Estimates of $T_c$ values fall in the shaded grey range, determined in Fig.~S1.
{\bf b} Schematic free-energy landscape for models whose change of activation energy is dominated by the change of the energy of local barriers (top), and for models where this change is dominated by the growth of cooperative effects over a distance $\xi$, corresponding to a distance in phase space $d_\xi$ (bottom).
{\bf c} Energy change along the minimum energy path associated with a local relaxation event of our considered model, at $T=0.47$. $\Delta r_{\rm path}$ indicates the norm of the displacement field along the path in the units of $\rho^{-1/3}$, with $\rho$ the number density. 
{\bf d} Trajectories of the particles with the largest displacement as the system moves along the minimum energy path with energy profile illustrated in {\bf c}. The size of each particle has been scaled down by a factor of four for visualization purposes.\label{Fig1:intro}}
\end{figure*}

This work focuses on the mechanism inducing the growth of the activation energy in liquids, which is the central problem of the glass transition.
For our purpose, it is useful to classify theories into two groups. 
In some popular approaches, 
the increase of the activation energy $E_a$  stems from the growth of a correlation length $\xi$, as sketched in Fig. \ref{Fig1:intro}b. 
In some views, including Adam-Gibbs and Random First Order Theory  \cite{Lubchenko01,Biroli12,tanaka2010critical}, $\xi$ characterizes a growing order associated with a thermodynamic phase transition. Below some intermediate temperatures where local `hopping processes' may affect the dynamics  \cite{gotze1988scaling,schweizer2003entropic,vollmayr2002dynamical,charbonneau2014hopping}, the slow process corresponds to breaking this order, which costs an energy $E_a\sim \xi^a$ where $a$ is some exponent. Kinetic constraints model (KCMs) \cite{keys2011excitations,ritort2003glassy} are other theories for which thermodynamics properties are instead trivial. 
Popular KCMs consider independent defects that cost an energy $U$, whose density decreases as $\exp(-U/T)$  under cooling, leading to a growing distance between defects $\xi$. For certain rules on the kinetic of defects, including the popular `East model' \cite{garrahan2002geometrical}, $E_a$ grows with $\xi$ and becomes much larger than the cost $U$ of moving a single defect.

By contrast, free volume \cite{turnbull1961free} or elastic \cite{dyre2006colloquium,rainone2020pinching,JeppeEdan} models assume that the activation energy is not cooperative as sketched in Fig.~\ref{Fig1:intro}b: it is governed by the energy barrier of elementary rearrangements, independently of $\xi$. In elastic models $E_a$ is controlled by the high-frequency shear modulus indeed known to grow in fragile liquids \cite{dyre2006colloquium} or alternatively by a local elastic modulus \cite{rainone2020pinching,JeppeEdan}. These views are consistent with the observation that thermal relaxation occurs more often in softer ~\cite{widmer2008irreversible} or more plastic regions  ~\cite{Li2022, Lerbinger2022}.
Obviously, the activation energy may also have a contribution from localized events and one associated with cooperative dynamics. In some scenarios, cooperative effects become dominant at low temperatures~\cite{Mirigian2013}.

Even though local and cooperative theories are based on differing physical assumptions, they both find empirical support at the level of correlations between observables they predict. 
Henceforth, our contention is that to distinguish these views, one must go beyond correlations and systematically measure the energy barrier of the localized activation events driving structural relaxation, illustrated in Fig.~\ref{Fig1:intro}c and d.

Previous numerical studies related the relaxation dynamics to the exploration of the potential energy landscape $E(\bf{r})$.
At low temperatures, this exploration proceeds via consecutive activated transitions, or excitations, through which the system transits from a  minimum of $E(\bf{r})$, called inherent structures (IS), to an adjacent one. 
Earlier works~\cite{Doliwa2003,Heuer08} focused on rather high temperatures and clarified that irreversibility occurs through a sequence of elementary excitations,  also evidenced experimentally \cite{Simmons2012,Cicerone2014,cicerone2023excitation}, corresponding to a displacement of a temperature-independent number of particles. 
Other studies considered the few lowest-energy excitations at lower temperatures to investigate the plastic and quantum properties of glassy solids \cite{Wencheng19, Schober93, Wang19, Khomenko20, Wencheng20}.  
Such low-energy excitations are well-known to be already active at temperatures much smaller than the glass transition. 
They thus cannot by themselves relax the material into a liquid. Overall, in the absence of a comprehensive description of how the distribution of energy barriers of localized relaxation events depends on temperature, it is hard to establish if the gradual increase in these energy barriers drives the formation of glasses, as suggested by local theories, or conversely, if the dynamics' slowdown results from the increasing spatial organization of these events, as posited by cooperative theories.

In this work, we first introduce SEER, an algorithm to systematically uncover high-energy excitations in numerical glasses. SEER allows us to measure the distribution $N(E_{\rm loc})$ of local energy barriers $E_{\rm loc}$ at various temperatures, around a given configuration. We use it in a system of continuously polydisperse particles. Such systems are known to reproduce the main facts of the glass transition, and can be efficiently thermalized \cite{ninarello2017models}, allowing us to explore a wide range of time scales.  
A remarkable result is that $N(E_{\rm loc})$ is essentially shifted to higher energy under cooling. 
This shift corresponds to an increase in local barriers, which must contribute to the rise in activation energy $E_a(T)$, alongside potential cooperative effects. 
Furthermore, we introduce a new empirical method based on flash reheating and cooling protocols, which may also be used experimentally to extract $E_a(T)$ without fitting parameters.
Remarkably, our observations show that the increase in activation energy resulting from these local barriers accurately and fully accounts for the growth of the relaxation time over seven orders of magnitude. Thus local effects, rather than cooperative ones, control the dynamics near the glass transition in this liquid. Our analysis can now be extended to any liquid model to measure the degree of cooperativity of their dynamics.

\begin{figure*}[!!ht]%
\centering
\includegraphics[width=0.9\textwidth]{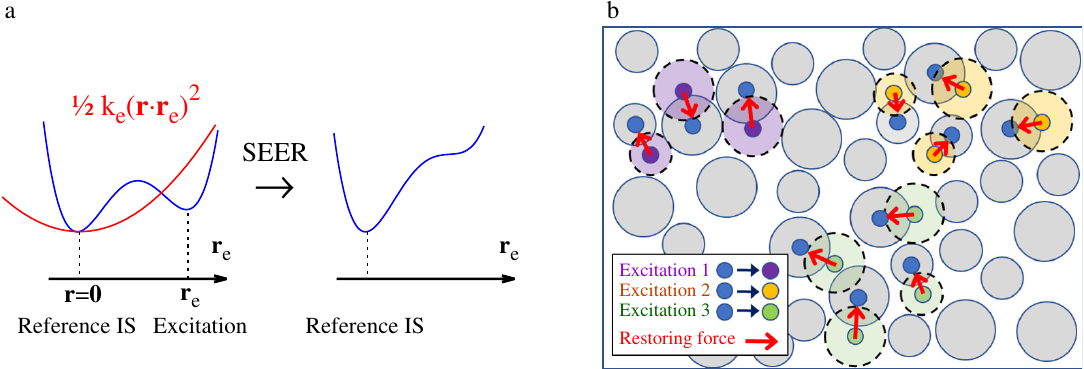}
\caption{
Schematic illustration of SEER's working principle. {\bf a} The algorithm suppresses a discovered excitation by transforming it into a saddle bifurcation; {\bf b} in real space, it corresponds to adding restoring forces opposite to displacements along the suppressed excitations. 
}\label{Fig1:seer_schematic}
\end{figure*}

\section*{SEER: Systematic Excitation ExtRaction}\label{subsec:seer}
We seek to uncover the excitations around a given inherent structure, which is a minimum of $E(\bf{r})$ obtained by quenching an equilibrated configuration, determining the number of excitations with activation energy $E_{\rm loc}$, $N(E_{\rm loc})$.
A standard method \cite{Doliwa2003, Wencheng22,Khomenko21} to uncover excitations consists in performing a small temperature cycle of duration $t_w$ at a temperature $T_w$ before descending again the energy, possibly toward a new minimum. 
$t_w$ and $T_w$ are chosen so that typically none or a single excitation is triggered. 
In cases where a few excitations are triggered simultaneously, algorithms were developed to disentangle them if they are spatially separated \cite{Wencheng20}, which we systematically use below. 
Unfortunately, such a method detects a given excitation with a probability $\propto t_w e^{-E_{\rm loc}/T_w}$:  the first few lowest-energy excitations are found extremely often, already at low $T_w$.  At such temperatures, higher energy ones are essentially never obtained. 

To overcome this limitation and access higher-energy excitation, we have developed a novel algorithm, SEER, whose detailed implementation is specified in \rev{SI}.
To access high-energy excitations, SEER modifies the energy functional once an excitation is found.
This modification is the minimal one that ensures that this excitation cannot be triggered again. 
It allows us to slowly increase the next cycling temperature $T_w$, so as to discover sequentially higher energy excitations starting from the same configuration.

Specifically, consider an inherent structure ${\bf r^*}$. To lighten notations, we consider ${\bf r^*}$ as specifying the reference position for all particles, such that ${\bf r}$ denotes the displacement field from that configuration. 
Once we discover an excitation via a temperature cycle as discussed above, i.e. a new minimum ${\bf r}_e \neq {\bf 0}$ of the potential energy, we modify the energy landscape along a single direction in phase space. 
Specifically, we add a term $\frac{1}{2} k_e ({\bf r} \cdot {\bf r}_e)^2$ to the Hamiltonian, which penalizes motion in the direction ${\bf r}_e$ of the discovered excitation.
We fix $k_e$ to the minimum value ensuring that $E({\bf r}) + \frac{1}{2} k_e ({\bf r}\cdot{\bf r}_e)^2$ has not a minimum near ${\bf r}_e$,
as illustrated in Fig.\ \ref{Fig1:seer_schematic}a,b. 
We repeat this procedure for each discovered excitation, such that future thermal cycles search excitations of the SEER energy:
\begin{equation}
    E_{\rm seer}({\mathbf r}) = E({\mathbf r}) + \sum_{e=1}^{N_e}\frac{1}{2} k_e ({\mathbf r}_e\cdot {\mathbf r})^2,
\end{equation}
where $N_e$ is the number of excitations already in our catalogue. 
We add to our catalogue the excitations of $E_{\rm seer}({\mathbf r})$ that are also excitations of $E$, and stop the run when $N_e$ reaches a prescribed value. 
We discuss the stability of SEER with respect to its control parameters in Fig.~S2.

Once an excitation is detected, we use the nudge-elastic-band (NEB) method~\cite{neb1} 
to determine the minimal energy path connecting the two inherent structures, as illustrated in Fig.~\ref{Fig1:intro}c,d. 
Obviously, this procedure is performed using the real energy $E({\bf r})$, in the absence of added potential. 
It yields both the barrier height $E_{\rm loc}$ and the number of intermediate maxima.

In principle, SEER's modification to the Hamiltonian may hinder the discovery of new excitations in the vicinity of previously discovered ones.  SEER alleviates this effect by choosing a minimal perturbation $k_e$ to the energy. 
More importantly, in a given run of SEER the order of apparition of excitations is stochastic. 
To estimate the number of excitations $N(E_{\rm loc})$ more accurately, we perform many runs, each delivering a catalogue of $N_e$ excitations.
We then merge these into a more complete library, as detailed in Fig.~S3.
By definition, $N(E_{\rm loc})$ counts the excitations obtained when the number of merged catalogues diverges. 
Practically, $N(E_{\rm loc})$ converges below some $E^{\rm conv}$ that depends on the number of merged catalogues. Below, we consider 50 catalogues. It will correspond to a number of excitations $N_e$ of the order of a hundred below $E^{\rm conv}$, implying that SEERS only constraints a tiny fraction of the degrees of freedom, $N_e/dN < 1\%$.

\section*{Choice of Model \label{sec:model}}
We consider the three-dimensional soft repulsive particles of Ref.~\cite{Lerner19}. This system belongs to the class of modern numerical models whose polydispersity allows us to use the efficient `swap' algorithm \cite{Glandt84,gutierrez2015static,Ninarello17,Brito18} to reach thermal equilibrium on a range of temperatures comparable to experiments. Such models  are receiving a considerable attention  \cite{ninarello2017models, Nishikawa2022, Berthier2023, Guiselin2022, Scalliet2022, berthier2019zero, Scalliet19, Ciarella2023}, as they capture the hallmarks of the glass transition.  
The size of the particles $\sigma$ is distributed as $p(\sigma) \propto \sigma^{-3}$ with $\sigma\in [\sigma_{\rm min}, 2.2\sigma_{\rm min}]$.  
Similar systems capture both the usual dynamical \cite{Ninarello17} and thermodynamical \cite{berthier2019configurational} properties of experimental liquids. 
The interaction is given by the purely repulsive potential 
\begin{equation}
V(r_{ij}) = \epsilon\left[ \left(\frac{\sigma_{ij}}{r_{ij}}\right)^{10} + \sum_{l=0}^{3} c_{2l}\left(\frac{r_{ij}}{\sigma_{ij}}\right)^{2l}\right]
\end{equation}
for $r_{ij}<x_c =1.4$ and vanishes at larger distances, with $c_{2l}$ chosen to enforce continuity at $x_c$ up to $3$ derivatives, and with an interaction range $\sigma_{ij}$ function of $\sigma_i$ and $\sigma_j$.   We study systems of $N=2000$ particles of mass $m$ at number density $\rho = 0.58$ in cubic simulation boxes with periodic boundary conditions via molecular dynamics (MD) simulations in the canonical ensemble. 
We express mass in units of $m$, temperature  in units of $\epsilon$,  lengths in units $\rho^{-1/3}$, and time in units of $\sqrt{m\sigma_{\rm min}^2/\epsilon}$.
Using an integration timestep $dt=0.01$, we simulate up to a maximum time of $10^7$, at the lowest considered temperature.

We  extract the relaxation time $\tau$ from the decay of the overlap function
\begin{equation}
Q(t) = \frac{1}{N} \Biggl \langle \sum_i^N e^{-\left(\frac{\Delta r_i(t)}{w}\right)^2} \Biggr \rangle
\end{equation}
averaged over 20 independent realizations, where $\Delta r_i(t)$ is the displacement of particle $i$ during a time interval $t$, and $w=0.3\rho^{-1/3}$.  
We define the relaxation time as $Q(\tau) = 1/e$.
\rev{We show in Fig. S4 that the self-scattering function evaluated at the first peak of the static structure or at smaller ones, and a collective overlap function, yield consistent estimations for the relaxation time.}

Using MD from equilibrated configurations, we can access the dynamics on about 6 decades. However, previous studies~\cite{Guiselin2022,Scalliet2022} have found in related models that the shape of the correlation function does not significantly depend on the temperature \cite{Berthier2020}, a property referred to as time-temperature superposition (TTS)  \cite{van1998time}.
We have verified in Fig.~S4 that this property holds in our model and use it to infer $\tau$ even when the correlation function does not fully decay in our observation window, leading to the nine decades of observation of Fig.\ref{Fig1:intro}a.
As has been reported in other highly-polydisperse ~\cite{Berthier2017}, in more conventional model systems~\cite{Das2022} or in experimental data~\cite{Mallamace2010}, the model seems to display Arrhenius-like behavior (corresponding to straight lines in the logarithmic representation of Fig.\ref{Fig1:intro}a) above the onset temperature $T_{\rm on}\simeq 0.7$ as well as at the lowest temperatures, with a super-Arrhenius behavior in between. Our analysis below (using reheating methods) will show that the activation energy is, in fact, not constant in any of the temperatures probed.

\section*{Results of SEER}

\begin{figure*}[!!ht]%
\centering
\includegraphics[width=0.9\textwidth]{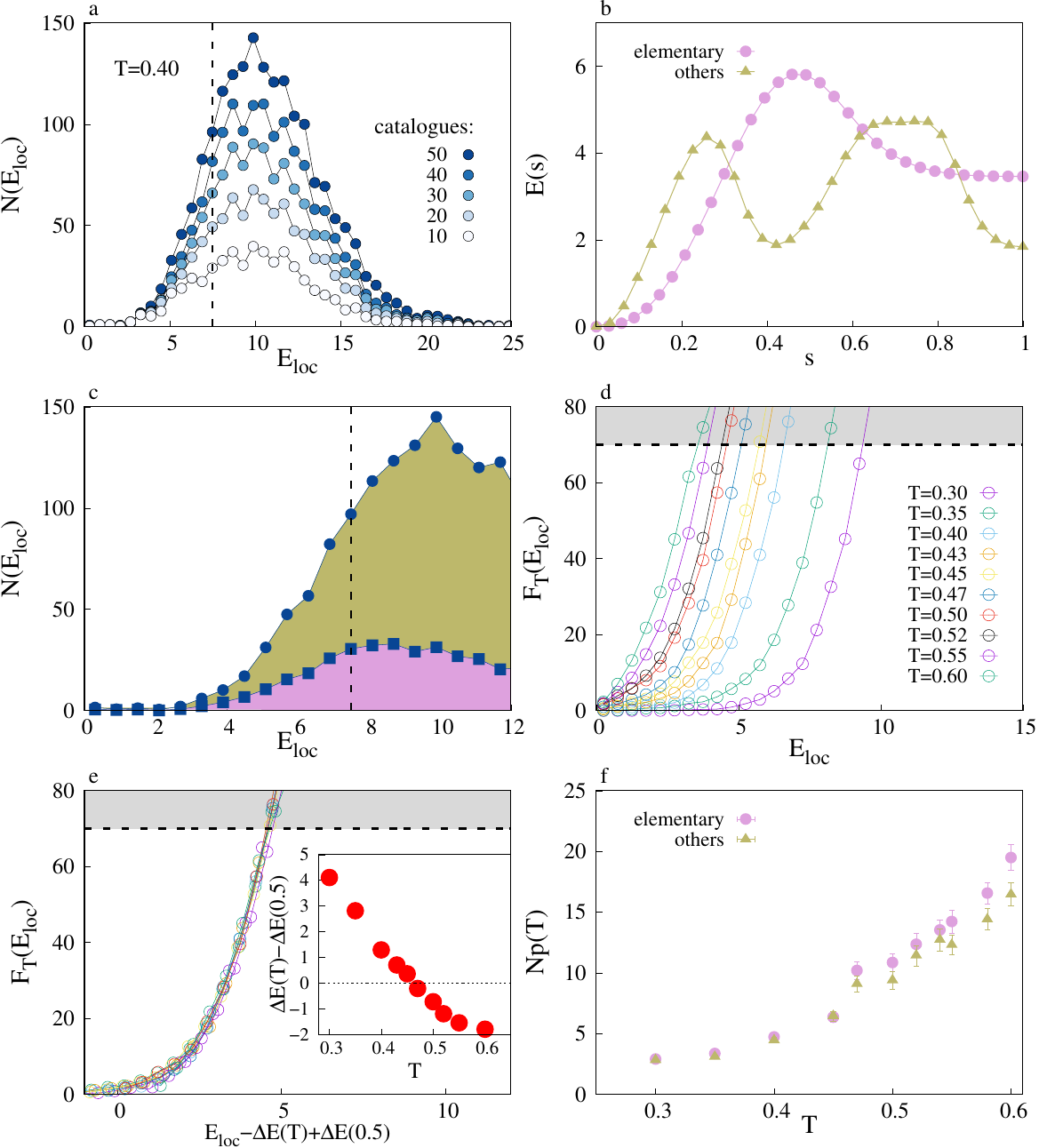}
\caption{
{\bf a} The number $N(E_{\rm loc})$ of distinct excitations with activation energy $E_{\rm loc}$ saturates with the number of considered catalogues, starting from small energies (see Fig.~S2 for further details).
$N(E_{\rm loc})$ increases by less than $5\%$ below an energy threshold (dashed line) when adding $10$ more catalogues. 
The figure refers to $T = 0.4$. 
{\bf b} The discovered excitations (circles) are simple (truly elementary) excitations with a single energy maximum (squares) along the minimum energy path or, instead, present multiple maxima. 
{\bf c} Below the energy convergence threshold, a large fraction of the discovered excitations are elementary.
{\bf d} The cumulative number of elementary excitations shifts to higher energies as the temperature decreases. 
At each $T$, we have at least $70$ elementary excitations with activation energy below the convergence threshold.
{\bf e} The collapse of the cumulative number of excitations on a master curve allows us to associate an energy shift $\Delta E(T)$ to each temperature.
(inset) Temperature dependence of the energy shift $\Delta E(T)$, using $T_c = 0.5$ as a reference.
{\bf f} Temperature dependence of the number of particles $Np$ involved in an excitation, averaged over all excitations with energy below the convergence threshold and on the elementary one with a single maximum. Excitations become increasingly localized as the temperature decreases.
In d-f, data are averaged over $10$ configurations with $N=2000$. In e,f $N_e = 50$.}\label{Fig1:seer}
\end{figure*}

\subsection{Temperature shifts the excitation density of state $N(E_{\rm loc})$}
Fig.~\ref{Fig1:seer}a illustrates $N(E_{\rm loc})$ for $T=0.4$ extracted from SEER, as a function of the number of catalogs used to build the library. After 50 runs $N(E_{\rm loc})$ approximately converges below some energy  $E^{\rm conv}$. We uncover about 200 excitations with $E_{\rm loc}<E^{\rm conv}$ (dashed line).
Many of these excitations have a single energy maximum along their reaction path, as illustrated in Fig.~\ref{Fig1:seer}b. The density $N(E_{\rm loc})$ of such single excitations is compared with the total density in Fig.~\ref{Fig1:seer}c. Considering all or only single excitations lead to similar results, in what follows we focus on the latter. 

To study the temperature dependence of excitations, we consider the cumulative number of elementary excitations  $F_T(E_{\rm loc})$ in Fig.~\ref{Fig1:seer}d. 
The figure reveals a depletion of low-energy excitations under cooling. 
Our key observation, illustrated in Fig.~\ref{Fig1:seer}e, is that different curves $F_T(E_{\rm loc})$ approximately collapse by simply shifting the energy by some value $\Delta E(T)$. 
\rev{Fig.~S5 shows that no collapse can be obtained by assuming that temperature scales the  energy axis by some factor.}
We show the temperature dependence for the energy shift $\Delta E(T)$  in the inset of Fig.~\ref{Fig1:seer}e. 
The error on this quantity is estimated in Fig.~S6.
The energy shift approaches a constant at temperatures close to the crossover one. 
It is consistent with the usual notion that above the crossover temperature $T_{\rm on} \simeq 0.7$, the properties of the inherent structures become temperature independent.
As the temperature decreases, $\Delta E(T)$ develops an approximate linear temperature dependence. 

Thus, the density of excitations evolves under cooling as if all excitations were increasing their energy by some amount $\Delta E(T)$. Making the assumption that the activation energy $E_a(T)$ is dominated by the magnitude of local barriers then implies that it must also shift by $\Delta E(T)$ under cooling. In other words, for $T_1,T_2$ at low temperatures we must have:
\begin{equation}
    E_a(T_1)-E_a(T_2)=\Delta E(T_1)-\Delta E(T_2)
    \label{pred}
\end{equation}
This is our main prediction, which we can test below without fitting parameters.

\subsection{Excitations become more localized under cooling\label{sec:localization}}
To show that individual excitations themselves do not become more and more cooperative under cooling, we estimate the number of particles they involve from their participation ratio. This number is $Np$, where $p = \frac{\left(\sum {\bf u}_i^2\right)^2}{N \sum {\bf u}_i^4}$ and \rev{${\bf u} = {\bf r_f}-{\bf r_i}$ the excitation displacement field, with {$\bf r_i$} and {$\bf r_f$} the initial and final particle configurations.} 
Fig.~\ref{Fig1:seer}f illustrates the temperature dependence of $Np$ averaged over all excitations below the energy convergence threshold.
Excitations involve a few particles and become increasingly localized at low temperatures.

The small number of particles involved in each excitation suggests that the properties of the latter do not depend on the system size.
Indeed, previous studies~\cite{Heuer08} demonstrated that the statistical features of the energy landscape of very small systems ($N < 100$) already represent the thermodynamic limit.

\section*{Local barriers control relaxation\label{subsec3}}
\begin{figure*}[t!]%
\centering
\includegraphics[width=0.8\textwidth]{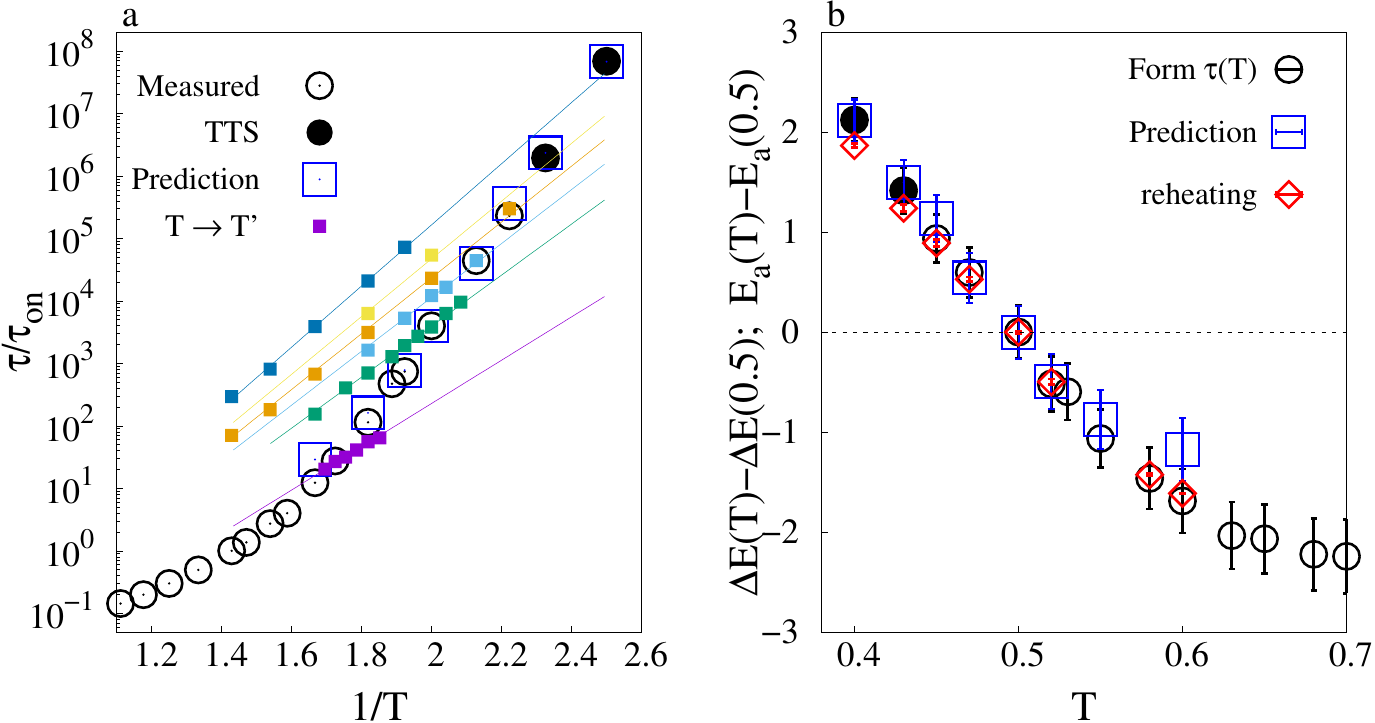}
\caption{{\bf a} Relaxation time as a function of the inverse temperature, as in Fig.\ref{Fig1:intro}a (circles).
Solid squares indicate the values $\tau(T \to T')$ of the relaxation time of systems equilibrated at temperature $T$ when the temperature is instantaneously changed to the value $T'$, \rev{for $T=0.40,0.43,0.45,0.47,0.50,0.58$}. 
At each $T$, lines are fit to $\tau(T \to T') = t_0 e^{E_a(T)/T'}$ on both reheating \rev{($T' \neq T$)} and equilibrated \rev{($T' = T$)} data so as to extract $t_0$ and $E_a(T)$. Squares and circles overlap when $T=T'$.
The value of $t_0$ has no clear $T$ dependence and, at each $T$, it slightly depends on the fitting range.
We adopt a conservative approach to account for this variability by fixing $t_0$ to an average value and using three standard deviations as an estimation of its error: $t_0\simeq (2.7\pm 1.3)10^{-5}\tau_{\rm on}$, where $\tau_{\rm on}$ is the relaxation time at the onset temperature. 
At $T=0.5$, we find $E_a(T=0.5) \simeq 9.4$ (other values of $E_a(T)$ can be extracted from panel b). Interestingly,  the fragility of this liquid is apparent, as the solid lines (corresponding to constant activation energy) increase systematically less fast than the $\tau(T)$ under cooling. Blue squares indicate our predictions $\tau_{\rm pred}(T)$  over almost five orders of magnitudes which follows from Eq.[\ref{pred}] and from our measurement of the shift in excitations' activation energy: $\tau_{\rm pred}(T) = t_0e^{(E_a(0.5)+\Delta E(T)-\Delta E(0.5))/T}$. 
{\bf b} Change in activation energy $E_a(T)-E_a(0.5)$ (circles) inferred from panel a. 
Errors are dominated by the uncertainty in $t_0$.
The change in activation energy is quantitatively predicted, without fitting parameters, by the excitations' activation energy shift, $\Delta E(T)-\Delta E(0.5)$ (squares). 
This result indicates that the change of activation energy is  governed by local barriers, instead of cooperative effects on a growing length $\xi$. Finally, if the activation energy is extracted from panel a using only the reheated data, one obtains a consistent estimation of $E_a(T)-E_a(0.5)$ (diamonds).
}
\label{Fig2:seer}
\end{figure*}

The prediction Eq.~[\ref{pred}] can be written alternatively as  $E_a(T) = E_0 + \Delta E(T)$, which implies for the relaxation time grows:
\begin{equation}
    \tau(T)=t_0e^{\frac{E_0 + \Delta E(T)}{T}}
    \label{pred2}
\end{equation}
In this expression, two parameters enter, $t_0$ and $E_0$. The latter can be readily removed to test our hypothesis by considering the change of activation energy with temperature as in Eq.~[\ref{pred}]. 
Yet, this quantity is interesting to estimate. 
The microscopic time $t_0$ depends on a diversity of factors, including the entropy of individual barriers~\cite{Pollak1998}, and no formal expression for it is available for the under-damped dynamics we use. It could depend on temperature. 
Yet if the change of dynamics is dominated by the energy change of local barriers, the evolution of $t_0$ can be neglected. Below we test such a model, where $t_0$ is fixed.

There is no established procedure to measure the activation energy $E_a$ and $t_0$.
Previous works, see e.g. ~\cite{Coslovich2018}, focused on the apparent activation energy $E_{\rm app} = \frac{d \log\tau}{dT} = E_a(T) - T\frac{d E_a}{dT}$. 
Unfortunately, it overestimates the actual activation energy unless $E_a$ is constant ~\cite{Struik1997}.
Regarding this point, note that while an Arrhenius-like behavior is often observed at low temperatures, it does not indicate a constant activation energy as often assumed, but rather an activation energy that depends linearly on temperature.
The apparent activation energy cannot distinguish these two scenarios.

We overcome these difficulties by introducing a novel empirical approach to determine $t_0$ and $E_a(T)$.
We consider equilibrated configurations at some temperature $T$ and suddenly change the temperature to some value $T'$. 
If $T'$ and $T$ do not greatly differ \footnote{If $T'$ is significantly larger than $T$, relaxation can occur via the nucleation of a hot liquid phase invading the material \cite{Berthier2020,Mehri2022}}, we expect the instantaneous relaxation time $\tau(T\to T')$ to follow an Arrhenius behavior in $T'$ with a constant activation energy $E_a(T)$: $\tau(T\to T') = t_0 e^{E_a(T)/T'}$. 
Fig.~\ref{Fig2:seer}a tests this expression successfully as both $T$ and $T'$ are varied. 
$t_0$ has no clear temperature dependence, and all curves are well-fitted by a single $t_0$ value. 
From these measurements, we extract $E_a(T)$.
\rev{Fig.~S7 provides more details on the reheating procedure.}

From these results, the hypothesis that the change of the activation energy under cooling is governed by the change of local barriers energy, formalized by Eq.~[\ref{pred}], can now be tested. 
Fig.~\ref{Fig2:seer}b compares these two changes, respectively $E_a(T)-E_a(T=0.5)$  and $\Delta E(T)-\Delta E(T=0.5)$, taking $T=0.5$ as a reference temperature. 
Remarkably, the agreement is quantitative. 
Equivalently stated, this result implies that knowing the relaxation time at some intermediate temperature, e.g. $T=0.5$, we can predict the dynamics at lower temperatures and thus the fragility of the system, as illustrated in Fig.~\ref{Fig2:seer}a. 
At very low temperatures where we cannot relax SWAP-equilibrated configurations via MD simulations, measuring the activation energy shift via SEER is still possible. We exploit this opportunity to estimate the relaxation time in this low-temperature regime in Fig.~S8.
\rev{
The rise in activation energy upon cooling characterizes the fragility of our model. Figure S8 illustrates that the activation energy grows by 50\% beyond its value at the onset temperature at $T=0.4$, and by 100\% at $T=0.3$.}

These results do not imply that jumping over a single barrier is generally sufficient to relax the system into a new state of similar energy. Excitations tend to be in their lower-energy state, thus performing a single jump most often increases energy, and can be followed by the reverse jump. Denote by $n_0$ the number of local barriers that must act sequentially in the same location to form an irreversible event, i.e. to find a new state of energy similar to the original one. $n_0$ characterizes the cooperativity of the dynamics. In theories such as RFOT or the East model, the growth of the activation energy $E_a$ under cooling is due to a growth of cooperativity $n_0$. Such an effect would contribute to the activation energy, in addition to the individual growth of local barriers we computed. This contribution is undetectable, thus the dependence $n_0$ on temperature can be neglected in the liquid considered. 

\section*{Conclusion}\label{sec13}
We have introduced SEER, a novel method to analyze the energy landscape of glasses, which systematically measures the elementary rearrangements around any given energy minimum. Our central result is that the density of state of these excitations is shifted toward higher energy under cooling. Using a novel method based on rapid temperature changes to analyze the dynamics and extract the activation energy,  we showed that this shift precisely predicts the liquid slowing down. This finding reveals that up to milliseconds time scales for which such  liquids can be simulated, fragility is  controlled by the change of energy of individual excitations, which shrink in size under cooling. It is not due to the fact that the dynamics becomes more and more cooperative.

At the practical level, the reheating methodology we introduced as well as SEER  have potential to be of broad applicability. Concerning the former, developing sufficiently fast experimental temperature jumps \cite{hecksher2019fast}  would be of high interest to measure directly the activation energy in super-cooled liquids.  Concerning SEER, it is more directly usable in numerical models where the energy landscape of inherent structures is indicative of the dynamics. It excludes hard sphere interactions or interactions with very narrow wells \cite{Eckert2002,Pham2002,Sciortino2002}, where the free energy landscape must be analyzed instead.  Implementing SEER would then require more involved methods, such as using effective potentials \cite{ brito2009geometric,degiuli2015theory,ghosh2010density,altieri2018higher,altieri2016jamming}.

We focused on modern numerical liquids made of poly-disperse particles, which are receiving a considerable attention \cite{ninarello2017models, Nishikawa2022, Berthier2023, Guiselin2022, Scalliet2022, berthier2019zero, Scalliet19, Ciarella2023} as they can be simulated on time scales closer to experimental conditions.  
It can be argued that these systems are specific, with small particles moving faster than larger ones~\cite{Pihlajamaa2023}. 
Yet, just as for multicomponent metallic glasses~\cite{Meyer2004,Voigtmann2009,Bartsch2010} or silica where some atoms move much faster than others, these models present the usual hallmarks of glassy dynamics~\cite{Berthier2017,Das2022}. 
More generally, the methods presented here will allow to test the degree of cooperativity of any liquid model. Particularly interesting cases are systems displaying a liquid-liquid phase transition \cite{sastry2003liquid} or hexatic ordering  \cite{tanaka2010critical}, in which cooperative effects may well influence the activation energy.

Conceptually, our results indicate that a dynamical correlation length $\xi$ can grow while not affecting the relationship between relaxation time and local barriers.  
Such a scenario is captured in simplified descriptions of interacting excitations called `elasto-plastic models', which are cellular automata initially introduced to study the plasticity of amorphous materials \cite{Nicolas2018}.  Their relaxation time is controlled,  as in our case,  by the density of excitations around visited configurations $N(E_{\rm loc})$ \cite{popovic2021thermally,ozawa2022elasticity}, which also presents a gap or depletion similar to our observations.
These models display a growing correlation length under cooling \cite{ozawa2022elasticity}. It can be explained in terms of slow avalanches of activated processes \rev{\cite{tahaei2023scaling, de2024dynamical}, whose statistics matches the longest available MD simulations of dynamical heterogeneities \cite{guiselin2021microscopic}, an agreement that holds for a broad range of kinetic rules \cite{gavazzoni2023testing}.} Thus, a description of the glass transition based on local barriers appears quantitatively consistent with both the slowing down of the dynamics presented here, and with the spatial correlations of the dynamics studied in these works. \rev{By themselves, the latter cannot distinguish different scenarios, as pointed out in \cite{bouchaud2024dynamics}.}

Finally, an important question for the future is what makes local barriers grow under cooling. Do they follow a high-frequency elastic modulus \cite{dyre2006colloquium}, or instead local (instead of global) elasticity~\cite{JeppeEdan}, perhaps affected by the presence of locally-favored structures ~\cite{coslovich2007understanding,royall2015role}? Alternatively, are they controlled by the varying geometry of elementary rearrangements under cooling? Some evidence suggests that these rearrangements are shaped by the presence of a dynamical transition at some $T_c$ \cite{Wencheng22}.  
The systematic extraction of excitations SEER allows may give a new handle to address this question in the future.

\begin{acknowledgments}	
We thank E. Lerner for providing some of the equilibrium configurations we investigated. We thank the Simons collaboration as well as L. Berthier, G. Biroli, C. Brito, C. Gavazzoni, E. Lerner,  M. Muller, M. Popovic, M. Ozawa and A. Tahaei for discussions.
M.P.C.  discloses support for the research of this work from Singapore Ministry of Education [MOE-T2EP50221-0016]. M.W acknowledges support from the Simons Foundation
Grant (No. 454953 Matthieu Wyart) and from the
SNSF under Grant No. 200021-165509.
\end{acknowledgments}


%
\end{document}


\title{Supplemenary Information for: The energy cost of local rearrangements, not cooperative effects, makes a model glass solid}

\author{Massimo Pica Ciamarra}\email{massimo@ntu.edu.sg}
\author{Wencheng Ji}
\author{Matthieu Wyart} \email{matthieu.wyart@epfl.ch}
\maketitle

\subsection*{SEER: Systematic Excitation ExtRaction~\label{sec:appendix_SEER}}
SEER constructs a catalogue of excitations associated with an inherent structure ${\bf r} = 0$ of a system governed by the energy functional $E$. 
To identify the excitations with the smallest activation energy, SEER searches excitations via thermal cycles performed at a temperature $T_w$, which is initially small $T_w\simeq 0.05$, and gradually increased by $\Delta T_w \simeq 0.005$. 
The velocities ${\bf v(t=0)}$ are initially drawn from the equilibrium distribution at $T_w=0.05$. This is the only source of randomness in the algorithm. When $T_w$ is changed, particle velocities are set by rescaling these initial velocities.

We detail SEER's working principle below:
\begin{enumerate}
    \item[1.] We set the temperature to $T_w$ by rescaling the initial velocities ${\bf v}$.
    \item[2.] We perform a thermal cycle evolving the system up to a maximum time $t_w$, while controlling the temperature via velocity rescaling. 
    Thermal cycles at temperature $T_w$ and duration $t_w$ uncover excitations with a typical maximal activation energy $\propto T_w \log(t_w/t_0)$. 
    As such, if the thermal cycle reaches the maximum allocated time without uncovering any excitation, we expand the activation energy observation window increasing $T_w$ by $\Delta T_w$ and move back to step 1. 
    On the contrary, if the thermal cycle detects an excitation, we perform the following steps.
    \item[3.] We employ the dichotomy method to determine the lowest temperature within the interval $[T_w-\Delta T_w:T_w]$ at which a thermal cycle detects an excitation. Temperature is always controlled by rescaling the velocities set in 1.
    By lowering $T_w$, we increase the probability we resolve an elementary transition.
    \item[4.] If the excitation is complex, we decompose it into elementary excitations {${\bf r}_e$} as in Ref.~\cite{Wencheng20}.
    \item[5.] If the excitations of $E_{\rm seer}$ are also excitations of $E$:
    \begin{itemize}
        \item study the excitation with the nudge-elastic-band and add it to the catalogue.
        \item for each excitation ${\bf r}_e$, we find the minimum $k_e$ that makes $E+\frac{1}{2}k_e ({\bf r}\cdot{\bf r}_e)$ loose this excitation: descending this modified energy from ${\bf r}_e$ goes back to ${\bf r}=0$. We determine $k_e$ with $5\%$ accuracy through the dichotomy method.
        \item We modify $E_{\rm seer}$ by adding the excitation suppressing terms $\frac{1}{2}k_e ({\mathbf r}\cdot{\mathbf r}_e)$.
    \end{itemize}
    \item[6.] We repeat the algorithm from 1 to 5 to search for a novel excitation.
\end{enumerate}
~\newpage
\begin{figure}[h!]
\centering
\includegraphics[width=0.4\textwidth]{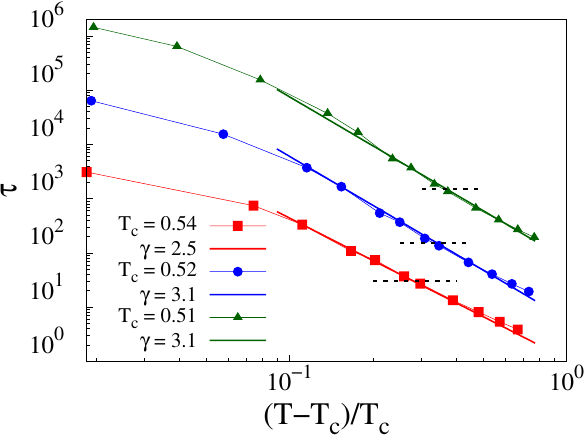}
\caption{
According to mode-coupling theory, the functional from $(T-T_c)^{-\gamma}$ should describe the relaxation time for $\tau > \tau_{\rm on}$.
We estimate $T_c$ as a useful reference by examining the behavior of the relaxation time for selected putative values of $T_c$ in Fig.~\ref{Fig:mct}.
For $T_c>0.54$, MCT only describes a small increase in relaxation time, up to $\tau \leq 10\tau_{\rm on}$. 
For $T_c < 0.51$, MCT underestimates the relaxation time in an increasingly large temperature window.
We thus estimate $T_c \in [0.51:0.54]$.
Data are shifted vertically for clarity. 
Dashed lines mark the value of the relaxation time at the onset temperature, $\tau_{\rm on}$. 
}
\label{Fig:mct}
\end{figure}
\newpage

\begin{figure}[h!]
\centering
\includegraphics[width=0.6\textwidth]{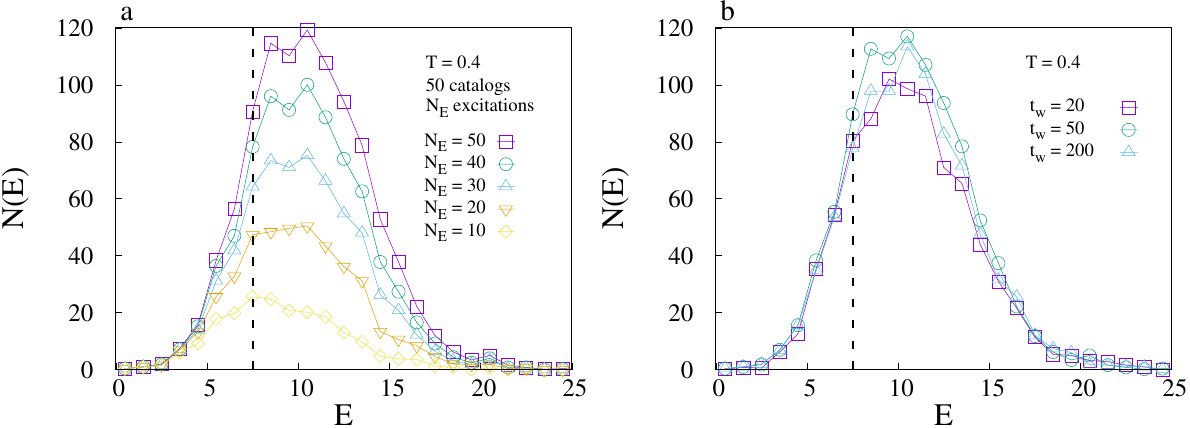}
\caption{
{\bf a} Activation energy dependence of the number of discovered excitations at $T=0.50$ using 50 catalogues, each containing $N_E$ excitations. Convergence shifts towards higher activation energies as $N_E$ increases.
{\bf b} In the energy convergence range, the number of discovered excitations $N(E)$ does not sensibly depend on the duration $t_w$ of the thermal cycles.
}
\end{figure}

~\newpage
\begin{figure}[h!]
\centering
\includegraphics[width=0.48\textwidth]{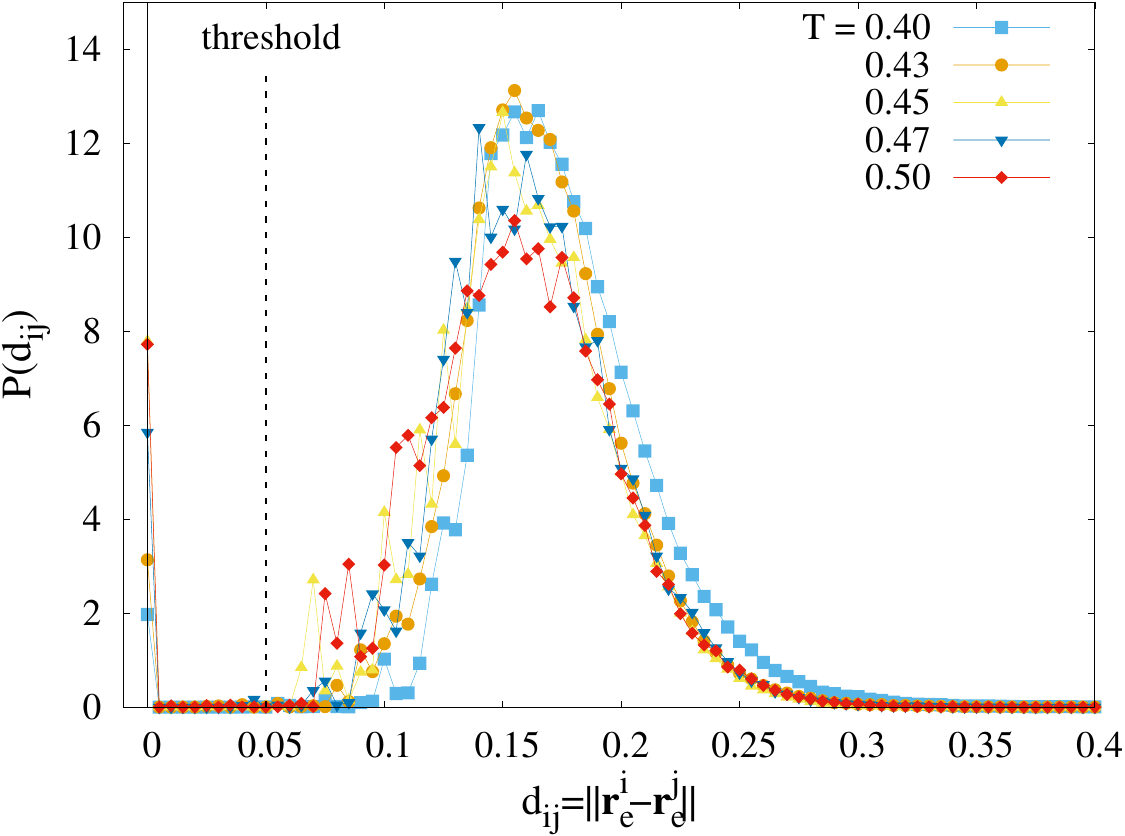}
\caption{
We associate many SEER's generated catalogues with the same initial configuration by changing the initial particle velocities.
Merging these catalogues yields our final catalogue.
When merging catalogues, we remove repeated excitations.
The distribution of the distance $d_{ij} =  \lVert {\bf r}_e^i-{\bf r}_e^j\rVert$ between discovered excitations illustrated in Fig.\ref{Fig:dist_excitations} reveals that two excitations are coincident, $d_{ij} \simeq 0$, or well separated. This finding allows us to robustly ensure our final catalogue only contains distinct excitations by using a threshold on the distance.
}
\label{Fig:dist_excitations}
\end{figure}

~\newpage
\begin{figure}[h!]
\centering
\includegraphics[width=0.6\textwidth]{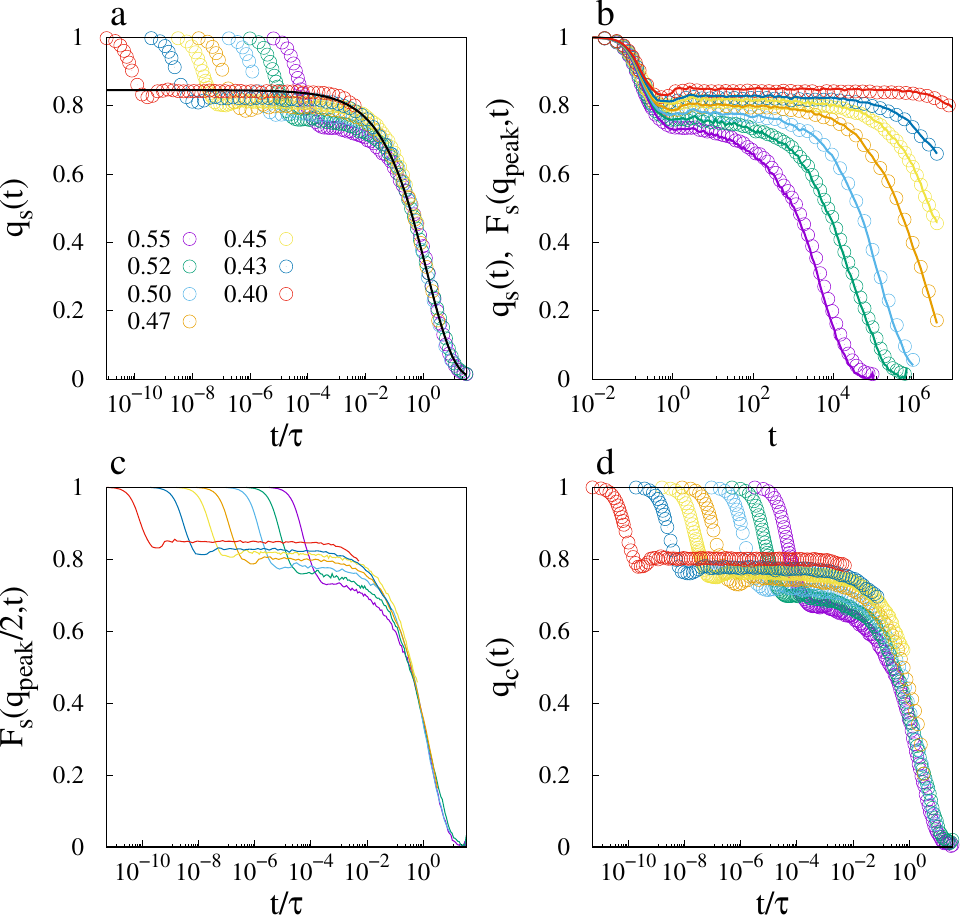}
\caption{{\bf a} The overlap correlation functions collapse on a stretched exponential function with stretching exponent $\beta \simeq 0.52$ (full line) when time is rescaled by the relaxation time. 
We use this property to estimate the relaxation time at the lowest temperatures, at which the overlap function does not fully decay within our simulation time.
{\bf b} The overlap correlation functions (symbols) behave essentially as the self-scattering function $F_s(q_{\rm peak},t) = \frac{1}{N} \sum_j e^{i {\bf q r_j}(t)} $ evaluated at a wavevector of the first peak of the static structure factor (full line). 
{\bf c} The self-scattering function evaluated at $q_{\rm peak}/2$ decays on the same time scale probed by the overlap.
{\bf d} The collective overlap decays on the same time scale probed by the self-overlap. We define the collective overlap as 
$q_{\rm c}(t) = \frac{o_{\rm c}(t)-o_{\rm c}^{\rm rand}}{o_{\rm c}(0)-o_{\rm c}^{\rm rand}}$, 
$o_{\rm c}(t) = \frac{1}{N}\sum_i \sum_j e^{-\frac{({\bf r}_i(t)-{\bf r}_j(0))^2}{w^2}}$, with $o_{\rm c}^{\rm rand} = \pi^{3/2}\rho w^3$ the expected value of $o_{\rm c}(t)$ in the absence of correlations between the two considered configurations.
}
\end{figure}

~\newpage
\begin{figure}[h!]
\centering
\includegraphics[width=0.6\textwidth]{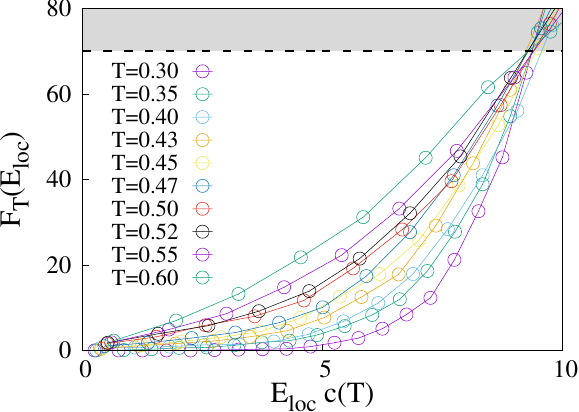}
\caption{In Fig~3e in the main text we have shown that the cumulative number of elementary excitations $F_T(E_{\rm loc})$ shifts to higher energy on cooling: $F_T(E_{\rm loc})$ data of different temperature collapse when plotted versus $E_{\rm loc}-\Delta E(T)$, with $\Delta E(T)$ a temperature-dependent energy shift. This figure shows that the $F_T(E_{\rm loc})$ cannot be collapsed when the energy axis is scaled by a factor $c(T)$. 
}
\end{figure}

~\newpage
\begin{figure}[t!]
\centering
\includegraphics[width=0.45\textwidth]{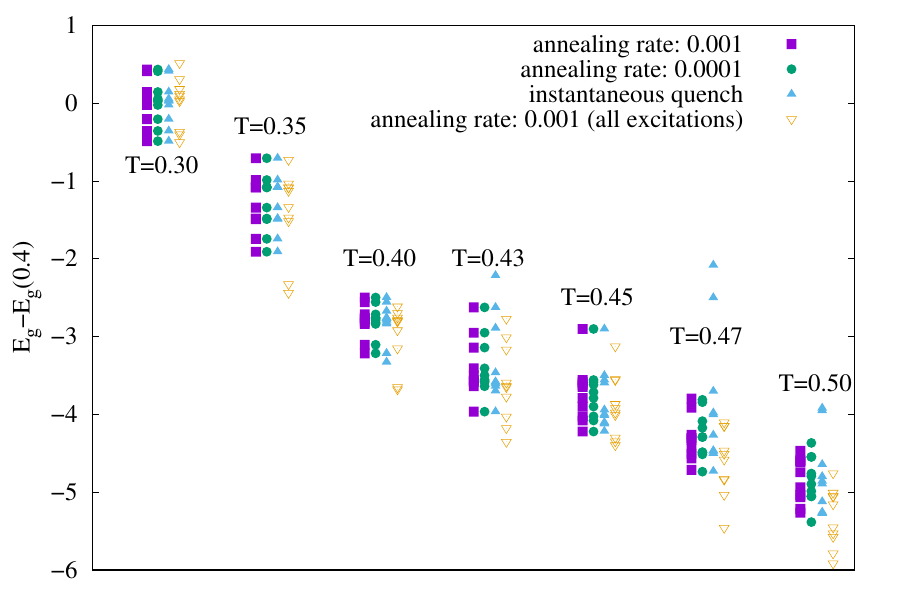}
\caption{
In the main text, we estimated excitations' energy shift by analysing the distribution of the energy barriers associated with the inherent structures of equilibrium configurations at each temperature.
We focused on data (i) obtained by quenching equilibrium configurations to a small temperature $T=0.3$, and then quenching to zero temperature at a rate $r = 10^{-3}$; (ii) averaged over ten independent samples; and (iii) restricted to elementary excitations with a single maximum. 
We illustrate the robustness of our result by comparing different approaches to estimate the energy shift $\Delta E$ of each configuration.\\
\\
Concerning (i), we avoid instantaneous quenches as it leads to a few outlier configurations with atypical $\Delta E$ at high $T$, as apparent in Fig.~\ref{fig:egcomapre}. Direct inspection indicated that these configurations present at least one excitation in its high-energy state, very close to a saddle transition (i.e. with a small barrier). This situation can lead to instabilities in SEER, if these excitations relax to low their low-energy state while other excitations are studied. Eliminating such configurations (not shown), or annealing the system, eliminates this problem and leads to consistent results independently of the annealing rate, as shown in Fig.~\ref{fig:egcomapre}.\\
\\
Concerning (ii), from the typical variance $\sigma^2$ of the ten data points at each temperature, we find our typical error in the estimation of $\Delta E$ to be $\Delta \Delta E\simeq \sigma/\sqrt{10} \simeq 0.1$.\\
\\
Concerning (iii), the analysis of all excitations (including complex ones, with multiple maxima as shown in open triangles), rather than the elementary ones, leads to analogous estimations of the energy shift at low temperatures, and slightly smaller estimations at the highest temperatures. This observation is plausibly related to the fact that, rigorously speaking, the density of states of all excitations is ill-defined. Indeed, in a large system, pairs of elementary excitations could be combined to generate a very large number of complex excitations. SEER seeks to avoid these fortuitous combinations (that can always occur by chance) by parsing complex excitations into elementary ones. Parsing is simple when two excitations are far away but difficult if they overlap. Overlaps will be more frequent at high temperatures, where excitations are more extended.
\label{fig:egcomapre}
}
\end{figure}

~\newpage
\begin{figure}[h!]
\centering
\includegraphics[width=0.90\textwidth]{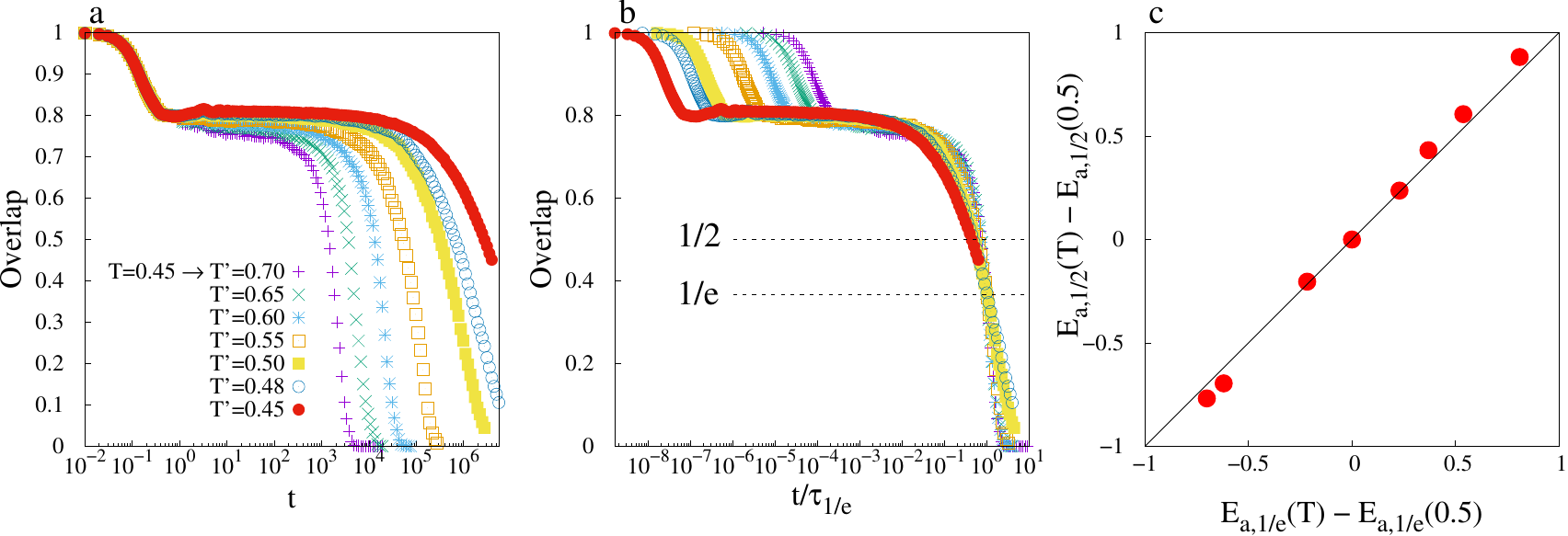}
\caption{
{\bf a} Overlap correlation functions under the re-heating dynamics. The temperature is changed from $T=0.45$ to $T'$ at $t = 0$. 
{\bf b} The overlap functions become increasingly stretched as $T'$ increases and hence do not collapse when time is scaled by the reheating relaxation time $\tau_{1/e}$ at which they reach $1/e$.
{\bf c} The Arrhenius dependence of the re-heating relaxation time on the re-heating temperature $T'$ allowed us to estimate the microscopic timescale $t_{0,1/e}$ and the activation energy $E_{a,1/e}$. Here, the subfix '$1/e$' indicates that the relaxation time is such that the overlap reaches $1/e$. To check the robustness of our approach, we repeated it by considering the system as relaxed when the overlap reaches $1/2 > 1/e$. This procedure allows us to estimate $t_{0,1/2}$ ($\simeq 0.7 t_{0,1/2}$) and $E_{a,1/2}(T)$.
We find $t_{0,1/2} \simeq 0.7 t_{0,1/e}$.
Critically, the figure shows that the change of activation energy with temperature is the same for $E_{a,1/e}(T)$ and $E_{a,1/2}(T)$.   
}
\end{figure}

~\newpage
\begin{figure}[h!]
\centering
\includegraphics[width=0.90\textwidth]{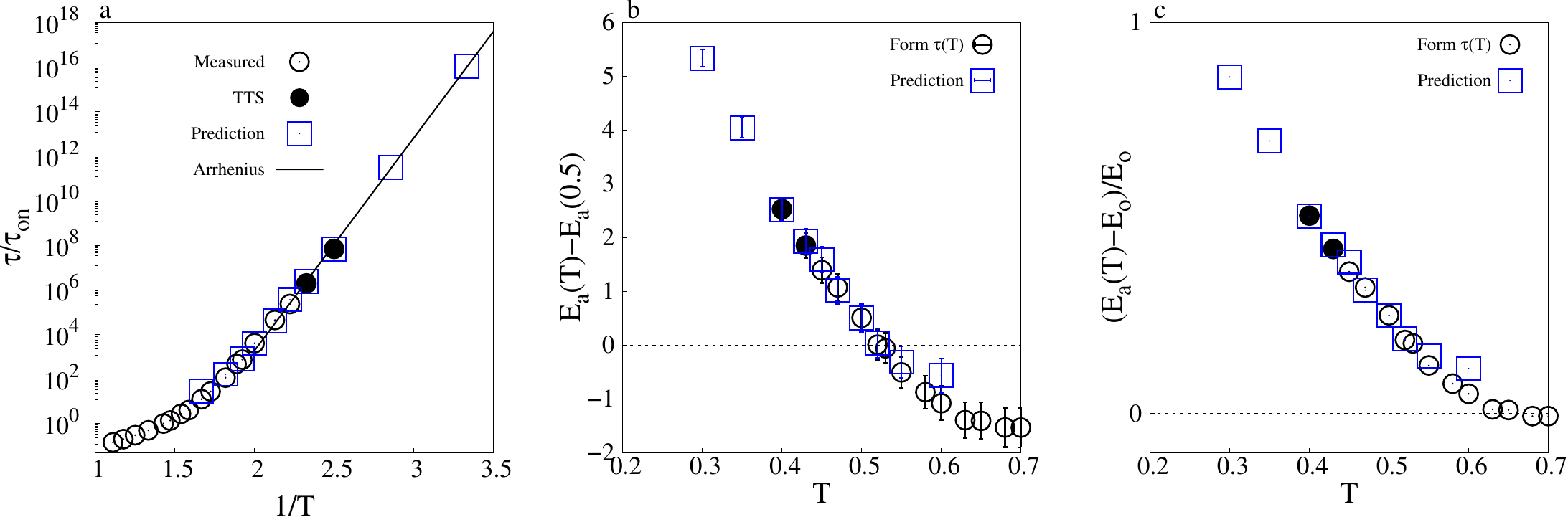}
\caption{
{\bf a,b} Inverse temperature dependence of the measured and estimated relaxation time and activation energies. 
The figure differs from Fig. 4 in the main text for the presence of data temperatures $T = 0.35$ and $0.30$.
We can measure the energy shift at these small temperatures by investigating swap-equilibrated configurations; 
On the contrary, their large relaxation time is inaccessible to numerical simulations.
These results are consistent with an effective Arrhenius behavior, which has been reported in some experimental data \cite{Mallamace2010}.
{\bf c} Change in activation energy relative to its estimated value at the onset temperature.  
\label{fig:verylowT}
}
\end{figure}

apsrev4-2.bst 2019-01-14 (MD) hand-edited version of apsrev4-1.bst
%
